\documentstyle[epsfig,amssymb,aps]{revtex}

\begin{document}
\author{L. B. Leinson}
\address{Institute of Terrestrial Magnetism, Ionosphere and Radio Wave Propagation\\
RAS, 142092 Troitsk, Moscow Region, Russia }
\author{A. P\'{e}rez}
\address{Departamento de F\'{\i }sica Te\'{o}rica Universidad de Valencia \\
46100 Burjassot (Valencia)\\
Spain}
\title{Relativistic approach to positronium levels in a strong magnetic field}
\maketitle

\begin{abstract}
We have investigated the bound states of an electron and positron in
superstrong magnetic fields typical for neutron stars. The complete
relativistic problem of positronium in a strong magnetic field has not been
succesfully solved up to now. In particular, we have studied the positronium
when it moves relativistically across the magnetic field. A number of
problems which deal with the pulsar magnetosphere, as well as the evolution
of protoneutron stars, could be considered as a field for application.
\end{abstract}


\section{Introduction}

Theoretical models for radio-pulsar emission mechanisms include the bound
states of a relativistic electron and positron in a superstrong magnetic
field. The year 1985 was marked by a number of papers, where the authors
tried to take into account the positronium contribution into the dispersion
equation of a photon which propagates in a superstrong magnetic field
(Leinson \& Oraevsky, \cite{LO85a}, \cite{LO85b}); Herold et al.,\cite{HRW85}%
; Shabad \& Usov, \cite{ShU85}). Nevertheless, the correct solution of the
completely relativistic problem for the bound states of an electron and
positron in a superstrong magnetic field remains unknown. Koller et al. \cite
{KMP88} have formulated the quadratic form of the completely relativistic
Dirac equation for bound states of an electron and positron in a magnetic
field of arbitrary intensity. However, they only found a solution for the
simplest case, when positronium does not move across the magnetic field.
There was also the work of Shabad and Usov \cite{ShU86}, where they tried to
solve the Bethe-Salpeter equation for a positronium atom relativistically
moving across the superstrong magnetic field. Unfortunately, they did not
take into account the retardation effect in the interaction between the
electron and positron. For this reason, we return to the problem of
positronium in a superstrong magnetic field. We suggest a solution of the
Bethe-Salpeter equation for positronium relativistically propagating across
the superstrong magnetic field. The results which contradict those given by
Shabad and Usov are discussed.

This paper is organized as follows. In Section 2 we state the problem by
considering the Bethe-Salpeter equation for a bound electron-positron pair
interacting with an external magnetic field. The problem can be simplified
when the adiabatic approximation (discussed in Section 3) holds, since in
this case only one Landau level for each particle has to be taken into
account. In section 4, we concentrate on the particular case when the
electron and positron occupy the same Landau level. As a particularly
important case, we devote Sect. 5 to the ground band of positronium levels,
and analyze some particular cases. In Appendix A we recall the explicit
formulae of the electron (or positron) wave function in the presence of an
external magnetic field.

\section{Bethe-Salpeter equation for a bound pair of electron and positron
in a strong magnetic field}

To find the bound states of electron and positron in a strong magnetic
field, we start from the Bethe-Salpeter equation for the scattering
amplitude, to the lowest order in the fine structure's constant $e^{2}\simeq
1/137$ 
\begin{equation}
\Gamma _{i_{j}}\left( p_{-},-p_{+}\right) =ie^{2}\int \frac{d^{3}{\bf p}%
_{-}^{^{\prime }}}{\left( 2\pi \right) ^{3}}\frac{d^{3}{\bf p}_{+}^{^{\prime
}}}{\left( 2\pi \right) ^{3}}\frac{d^{4}q}{\left( 2\pi \right) ^{4}}%
G_{il}\left( p_{-},p_{-}^{^{\prime }}\right) \gamma _{ls}^{\mu }\Gamma
_{st}(p_{-}^{^{\prime }}-q,-p_{+}^{^{\prime }}-q)\gamma _{tq}^{\nu
}G_{qj}\left( -p_{+}^{^{\prime }},-p_{+}\right) D_{\mu \nu }\left( q\right) 
\label{B_S}
\end{equation}
where $\widehat{G}\left( p_{-},p_{-}^{^{\prime }}\right) $ and $\widehat{G}%
\left( -p_{+}^{\prime },-p_{+}\right) $ are the propagators of the electron
and positron, respectively, in an external uniform magnetic field, which
will be taken as directed along the $Z$ axis. Here and henceforth, a $-$ ($+$%
) subscript denotes the electron (positron), and summation over repeated
indices will be understood. $D_{\mu \nu }\left( q\right) $ , with $q=(\omega
,{\bf q)}$ is the photon propagator. In Eq. (\ref{B_S}) $%
p_{-},p_{+},q,p_{+}^{^{\prime }}$ and $p_{-}^{^{\prime }}$ are the
four-momenta of the interacting particles.

It is convenient to study this equation in the mixed (coordinates-energy)
representation. To this purpose, we define the new unknown function 
\begin{equation}
\chi _{ij}\left( {\bf r}_{-},{\bf r}_{+};{\bf \varepsilon }_{-},{\bf %
-\varepsilon }_{+}\right) \equiv \int \frac{d^{3}{\bf p}_{-}}{\left( 2\pi
\right) ^{3}}\frac{d^{3}{\bf p}_{+}}{\left( 2\pi \right) ^{3}}{\bf \exp \ }%
\left( i{\bf p}_{-}{\bf r}_{-}\right) \Gamma _{ij}(p_{-},-p_{+}){\bf \exp \ }%
\left( i{\bf p}_{+}{\bf r}_{+}\right) .
\end{equation}
which depend on both individual space coordinates ${\bf r}_{\mp }$ and
energies of the electron and the positron. Their energies can be written as: 
\begin{equation}
{\bf \varepsilon }_{-}=\frac{E}{2}+\varepsilon ,\hspace{1in}{\bf \varepsilon 
}_{+}=\frac{E}{2}-\varepsilon 
\end{equation}
where the total energy $E$ of the bound pair is an integral of motion in the
stationary state we are considering. By the use of Fourier transformation of
the Green's functions 
\begin{equation}
\widehat{G}\left( p_{-},p_{-}^{^{\prime }}\right) =\int d^{3}{\bf r}%
_{-}^{^{\prime }}d^{3}{\bf r}_{-}\widehat{G}({\bf r}_{-}{\bf ,r}%
_{-}^{^{\prime }}{\bf \ ;\varepsilon }_{-}{\bf )\exp \ }[-i{\bf p}_{-}{\bf r}%
_{-}+i{\bf p}_{-}^{\prime }{\bf r}_{-}^{\prime }]
\end{equation}
\begin{equation}
\widehat{G}\left( -p_{+}^{\prime },-p_{+}\right) =\int d^{3}{\bf r}%
_{+}^{^{\prime }}d^{3}{\bf r}_{+}\widehat{G}({\bf r}_{+}^{\prime }{\bf ,r}%
_{+}{\bf \ ;-\varepsilon }_{+}{\bf )\exp \ }[i{\bf p}_{+}^{\prime }{\bf r}%
_{+}^{\prime }-i{\bf p}_{+}{\bf r}_{+}]
\end{equation}
we obtain the following equation: 
\begin{eqnarray}
&&\chi _{ij}\left( {\bf r}_{-},{\bf r}_{+};\frac{E}{2}+\varepsilon ,{\bf -}%
\frac{E}{2}+\varepsilon \right) =ie^{2}\int d^{3}{\bf r}_{-}^{^{\prime
}}d^{3}{\bf r}_{+}^{^{\prime }}G_{il}({\bf r}_{-}{\bf ,r}_{-}^{^{\prime }}%
{\bf ;}\frac{E}{2}+\varepsilon {\bf )}G_{qj}({\bf r}_{+}^{\prime }{\bf ,r}%
_{+}{\bf ;-}\frac{E}{2}+\varepsilon {\bf )}  \nonumber \\
&&\times \int \frac{d^{3}{\bf q}d\omega }{\left( 2\pi \right) ^{4}}D_{\mu
\nu }\left( {\bf \ \omega ,q}\right) {\bf \exp \ }\left[ i{\bf q}\left( {\bf %
r}_{-}^{\prime }-{\bf r}_{+}^{\prime }\right) \right] \gamma _{ls}^{\mu
}\chi _{st}\left( {\bf r}_{-}^{\prime },{\bf r}_{+}^{\prime };\frac{E}{2}%
+\varepsilon -\omega ,{\bf -}\frac{E}{2}+\varepsilon -\omega \right) \gamma
_{tq}^{\nu }
\end{eqnarray}
We choose the following gauge : 
\begin{equation}
A_{0}=0,{\bf A}=(0,Bx,0)  \label{gauge}
\end{equation}
so that $B_{x}=B_{y}=0$, $B_{z}=B$. In this way, the Green's function of the
electron takes the general form : 
\begin{equation}
G({\bf r}_{-}{\bf ,r}_{-}^{^{\prime }}{\bf ;}\varepsilon _{-}{\bf )}%
=\sum_{n\sigma }\int \frac{dk_{3}^{-}}{2\pi }\frac{dk_{2}^{-}}{2\pi }\exp
[ik_{3}^{-}(z_{-}-z_{-}^{^{\prime }})+ik_{2}^{-}(y_{-}-y_{-}^{^{\prime }})]%
\frac{\Psi _{n\sigma }^{\left( +\right) }(x_{-}+\frac{k_{2}^{-}}{eB}%
,k_{3}^{-})\overline{\Psi }_{n\sigma }^{(+)}(x_{-}^{^{\prime }}+\frac{%
k_{2}^{-}}{eB},k_{3}^{-})}{[\varepsilon _{-}-E_{n}(k_{3}^{-})+i0]},
\end{equation}

where $E_{n}(k_{3}^{-})=\sqrt{m_{n}^{2}+\left( k_{3}^{-}\right) ^{2}}$ and $%
m_{n}\equiv m\sqrt{1+2nb}$, with $b\equiv B/B_{0}.$ ($B_{0}=m^{2}/e=4.4%
\times 10^{13}G$ is the so-called {\em critical magnetic field}). The
electron wave functions $\Psi _{n\sigma }^{\left( +\right) }$ are bispinors
of positive frequency which correspond to Landau states of the electron in
the magnetic field. Each Landau state of the electron is marked by the
number $n=0,1,...$ which characterizes the quantized motion across the
magnetic field; $k_{3\text{ }}^{-}$ is the electron momentum along the
magnetic field; $\sigma =\pm 1$ corresponds to the spin projection of the
electron $s_{3}=\pm 1/2$. For many applications, it is convenient to use
wave functions of negative frequency for the states of positron (See
Appendix A). Then, the positron propagator has the following form: 
\begin{equation}
G({\bf r}_{+}^{^{\prime }}{\bf ,r}_{+}{\bf ;}\varepsilon _{+}{\bf )}%
=\sum_{n^{\prime }\sigma }\int \frac{dk_{3}^{+}}{2\pi }\frac{dk_{2}^{+}}{%
2\pi }\exp [-ik_{3}^{+}(z_{+}^{^{\prime }}-z_{+})-ik_{2}^{+}(y_{+}^{^{\prime
}}-y_{+})]\frac{\Psi _{n^{\prime }\sigma }^{\left( -\right)
}(x_{+}^{^{\prime }}-\frac{k_{2}^{+}}{eB},-k_{3}^{+})\overline{\Psi }%
_{n^{\prime }\sigma }^{(-)}(x_{+}-\frac{k_{2}^{+}}{eB},-k_{3}^{+})}{%
[\varepsilon _{+}+E_{n^{\prime }}(k_{3}^{+})-i0]}{},
\end{equation}

where $E_{n^{\prime }}(k_{3}^{+})=\sqrt{m_{n^{\prime }}^{2}+\left(
k_{3}^{+}\right) ^{2}}$. We assume that the center of mass of the
positronium can move relativistically in any direction with respect to the
magnetic field, i.e., we make no assumption about the center of mass
momentum 
\begin{equation}
P_{3}=k_{3}^{-}+k_{3}^{+}
\end{equation}
along the magnetic field, but we assume that the longitudinal relative
motion of the electron and positron is nonrelativistic. Therefore, the
relative momentum projection 
\begin{equation}
k_{3}=\frac{m_{n^{^{\prime }}}k_{3}^{-}-m_{n}k_{3}^{+}}{m_{n}+m_{n^{^{\prime
}}}}
\end{equation}
is much smaller than the electron mass. Let us write the electron and
positron momenta in the following form : 
\begin{equation}
k_{3}^{-}=\frac{m_{n}}{m_{n}+m_{n^{^{\prime }}}}P_{3}+k_{3}
\end{equation}
\begin{equation}
k_{3}^{+}=\frac{m_{n^{\prime }}}{m_{n}+m_{n^{^{\prime }}}}P_{3}-k_{3}
\end{equation}
Thus, we may expand the Landau levels energy as a series of $k_{3}$ : 
\begin{equation}
E_{n}(k_{3}^{-})\simeq \frac{m_{n}\sqrt{M_{nn^{\prime }}^{2}+P_{3}^{2}}}{%
M_{nn^{\prime }}}+\frac{P_{3}k_{3}}{\sqrt{M_{nn^{\prime }}^{2}+P_{3}^{2}}}+%
\frac{M_{nn^{\prime }}^{3}}{\left( M_{nn^{\prime }}^{2}+P_{3}^{2}\right)
^{3/2}}\frac{k_{3}^{2}}{2m_{n}}  \label{eland1}
\end{equation}
\begin{equation}
E_{n^{\prime }}(k_{3}^{+})\simeq \frac{m_{n^{\prime }}\sqrt{M_{nn^{\prime
}}^{2}+P_{3}^{2}}}{M_{nn^{\prime }}}-\frac{P_{3}k_{3}}{\sqrt{M_{nn^{\prime
}}^{2}+P_{3}^{2}}}+\frac{M_{nn^{\prime }}^{3}}{\left( M_{nn^{\prime
}}^{2}+P_{3}^{2}\right) ^{3/2}}\frac{k_{3}^{2}}{2m_{n^{\prime }}}
\label{eland2}
\end{equation}

where  $M_{nn^{\prime }}\equiv m_{n}+m_{n^{^{\prime }}}$. This yields 
\begin{equation}
E_{n}(k_{3}^{-})+E_{n^{\prime }}(k_{3}^{+})\simeq \sqrt{M_{nn^{\prime
}}^{2}+P_{3}^{2}}+\frac{k_{3}^{2}}{2\mu _{nn^{\prime }}}  \label{En}
\end{equation}
with the reduced mass 
\begin{equation}
\mu _{nn^{\prime }}=\frac{m_{n}m_{n^{\prime }}}{M_{nn^{\prime }}}\frac{%
\left( M_{nn^{\prime }}^{2}+P_{3}^{2}\right) ^{3/2}}{M_{nn^{\prime }}^{3}}
\end{equation}

\section{Adiabatic approximation}

In a sufficiently strong magnetic field $B>>10^{9}G$ the Larmor radius 
\begin{equation}
a_{L}=\frac{1}{\sqrt{eB}}
\end{equation}
is small with respect to the Bohr's radius 
\begin{equation}
a_{B}=\frac{1}{me^{2}}
\end{equation}
because : 
\begin{equation}
\frac{a_{B}}{a_{L}}=\sqrt{eB}\frac{1}{me^{2}}=\frac{\sqrt{b}}{e^{2}}>>1.
\label{11}
\end{equation}
When inequality (\ref{11}) holds, the Coulomb's binding energy $|\varepsilon
|\sim me^{4}$ is much smaller than the distance between Landau levels of the
electron (or positron). This means that the energies $\varepsilon _{\mp }$
of the bound electron and positron vary in a small vicinity of the Landau
levels they occupy. In this case, the poles near these Landau levels give
the principal contribution to the Green's functions of the electron and the
positron. Therefore, we can write : 
\begin{equation}
\hat{G}^{n}({\bf r}_{-}{\bf ,r}_{-}^{^{\prime }}{\bf ;}\varepsilon _{-}{\bf )%
}=\sum_{\sigma }\int \frac{dk_{3}^{-}}{2\pi }\frac{dk_{2}^{-}}{2\pi }\exp
[ik_{3}^{-}(z_{-}-z_{-}^{^{\prime }})+ik_{2}^{-}(y_{-}-y_{-}^{^{\prime }})]%
\frac{\Psi _{n\sigma }^{\left( +\right) }(x_{-}+\frac{k_{2}^{-}}{eB}%
,k_{3}^{-})\overline{\Psi }_{n\sigma }^{(+)}(x_{-}^{^{\prime }}+\frac{%
k_{2}^{-}}{eB},k_{3}^{-})}{[\varepsilon _{-}-E_{n}(k_{3}^{-})+i0]}
\end{equation}
\begin{equation}
\hat{G}^{n^{\prime }}({\bf r}_{+}^{^{\prime }}{\bf ,r}_{+}{\bf ;\varepsilon }%
_{+}{\bf )}=\sum_{\sigma }\int \frac{dk_{3}^{+}}{2\pi }\frac{dk_{2}^{+}}{%
2\pi }\exp [-ik_{3}^{+}(z_{+}^{^{\prime }}-z_{+})-ik_{2}^{+}(y_{+}^{^{\prime
}}-y_{+})]\frac{\Psi _{n^{\prime }\sigma }^{\left( -\right)
}(x_{+}^{^{\prime }}-\frac{k_{2}^{+}}{eB},-k_{3}^{+})\overline{\Psi }%
_{n^{\prime }\sigma }^{(-)}(x_{+}-\frac{k_{2}^{+}}{eB},-k_{3}^{+})}{%
[\varepsilon _{+}+E_{n^{\prime }}(k_{3}^{+})-i0]}{}
\end{equation}

Generally speaking, in the case of $n\neq n^{\prime }$ two terms
corresponding to the same total energy $E=\varepsilon _{+}+\varepsilon _{-}$
give the major contribution to the product of Green functions in the
right-hand side of the Bethe-Salpeter equation. The first of them
corresponds to an electron occupying the Landau level \ $n$, while the
positron occupies the level $n^{^{\prime }}$. The second term is the state
with the same energy, but with the electron occupying the level $n^{^{\prime
}}$ and the positron in the state $n$. Within this adiabatic approximation,
it is possible to characterize the positronium states in a strong magnetic
field by the almost-good quantum numbers $n$ and $n^{^{\prime }}$ , and by
other additional constants, which characterize the motion of the center of
mass and the relative motion of the electron and positron along the magnetic
field .

\section{Equation for the bound state wave function for equal Landau numbers}

If both the electron and the positron occupy the same Landau level, then
virtual photons with small values of $\omega \sim me^{4}\ll \left| {\bf q}%
\right| \sim me^{2}$ give the major contribution to the $e^{+}e^{-}$
interaction. Therefore, we can neglect $\omega $ in the photon propagator

\begin{equation}
D_{\mu \nu }\left( {\bf \ \omega ,q}\right) =g_{\mu \nu }\frac{4\pi }{\omega
^{2}-{\bf q}^{2}}\simeq -g_{\mu \nu }\frac{4\pi }{{\bf q}^{2}}
\end{equation}
This yields, for $n=n^{\prime }$: 
\begin{eqnarray}
&&\chi _{ij}\left( {\bf r}_{-},{\bf r}_{+};\frac{E}{2}+\varepsilon ,{\bf -}%
\frac{E}{2}+\varepsilon \right) =-ie^{2}\int d^{3}r_{-}^{^{\prime
}}d^{3}r_{+}^{^{\prime }}G_{il}^{n}({\bf r}_{-}{\bf ,r}_{-}^{^{\prime }}{\bf %
;}\frac{E}{2}+\varepsilon {\bf )}G_{qj}^{n}({\bf r}_{+}^{\prime }{\bf ,r}_{+}%
{\bf ;-}\frac{E}{2}+\varepsilon {\bf )}  \nonumber \\
&&{\bf \times }\int \frac{d^{3}q}{\left( 2\pi \right) ^{3}}\frac{4\pi }{{\bf %
q}^{2}}{\bf \exp \ }\left[ i{\bf q}\left( {\bf r}_{-}^{\prime }-{\bf r}%
_{+}^{\prime }\right) \right] g_{\mu \nu }\gamma _{ls}^{\mu }\int \frac{%
d\omega }{2\pi }\chi _{st}\left( {\bf r}_{-}^{\prime },{\bf r}_{+}^{\prime };%
\frac{E}{2}+\omega ,{\bf -}\frac{E}{2}+\omega \right) \gamma _{tq}^{\nu }
\end{eqnarray}
By performing integration over $d\varepsilon /2\pi $, we obtain the wave
function of the bound pair 
\begin{equation}
\hat{\Phi}({\bf r}_{-},{\bf r}_{+};E)=\int \frac{d\varepsilon }{2\pi }\chi
_{ij}\left( {\bf r}_{-},{\bf r}_{+};\frac{E}{2}+\varepsilon ,{\bf \ -}\frac{E%
}{2}+\varepsilon \right) 
\end{equation}
which verifies the following equation: 
\begin{equation}
\hat{\Phi}\left( {\bf r}_{-},{\bf r}_{+};E\right) =-ie^{2}\int \frac{%
d^{3}r_{-}^{^{\prime }}d^{3}r_{+}^{^{\prime }}}{\left| {\bf r}_{-}^{\prime }-%
{\bf r}_{+}^{\prime }\right| }\frac{d\varepsilon }{2\pi }\hat{G}^{n}({\bf r}%
_{-}{\bf ,r}_{-}^{^{\prime }}{\bf ;}\frac{E}{2}+\varepsilon {\bf )}\gamma
^{\mu }\hat{\Phi}\left( {\bf r}_{-}^{\prime },{\bf r}_{+}^{\prime };E\right)
\gamma _{\mu }\hat{G}^{n}({\bf r}_{+}^{\prime }{\bf ,r}_{+}{\bf ;-}\frac{E}{2%
}+\varepsilon {\bf )}  \label{wfe}
\end{equation}
Integration over $d\varepsilon ^{^{\prime }}/2\pi $ in the right-hand side
can be done with the help of the following formula :

\begin{equation}
\int_{-\infty }^{\infty }\frac{d\varepsilon }{2\pi }\frac{1}{\left(
\varepsilon +E/2-E_{n}(k_{3}^{-})+i0\right) \left( \varepsilon
-E/2+E_{n^{\prime }}(k_{3}^{+})-i0\right) }=\frac{i}{%
E-E_{n}(k_{3}^{-})-E_{n^{\prime }}(k_{3}^{+})}
\end{equation}
Thus, one has 
\begin{eqnarray}
Q_{ilqj}^{nn^{\prime }} &\equiv &\int \frac{d\varepsilon ^{^{\prime }}}{2\pi 
}G_{il}^{n}({\bf r}_{-}{\bf ,r}_{-}^{^{\prime }}{\bf ;}\frac{E}{2}+{\bf %
\varepsilon }^{^{\prime }}{\bf )}G_{qj}^{n^{\prime }}({\bf r}_{+}{\bf ,r}%
_{+}^{^{\prime }}{\bf ;-}\frac{E}{2}+{\bf \varepsilon }^{^{\prime }}{\bf )}%
=\sum_{\sigma }\sum_{\sigma ^{^{\prime }}}i\int \frac{dk_{3}^{-}}{2\pi }%
\frac{dk_{2}^{-}}{2\pi }\frac{dk_{3}^{+}}{2\pi }\frac{dk_{2}^{+}}{2\pi } 
\nonumber \\
&&\times {}\frac{\exp [ik_{3}^{-}(z_{-}-z_{-}^{^{^{\prime
}}})+ik_{2}^{-}(y_{-}-y_{-}^{^{\prime }})-ik_{3}^{+}(z_{+}^{^{\prime
}}-z_{+})-ik_{2}^{+}(y_{+}^{^{\prime }}-y_{+})]}{E-E_{n}(k_{3}^{-})-E_{n^{%
\prime }}(k_{3}^{+})}{}  \nonumber \\
&&\ \ \times \Psi _{n\sigma }^{i\left( +\right) }(x_{-}+\frac{k_{2}^{-}}{eB})%
\overline{\Psi }_{n\sigma }^{l(+)}(x_{-}^{^{\prime }}+\frac{k_{2}^{-}}{eB}%
)\Psi _{n^{^{\prime }}\sigma ^{^{\prime }}}^{q\left( -\right)
}(x_{+}^{^{\prime }}-\frac{k_{2}^{+}}{eB})\overline{\Psi }_{n^{^{\prime
}}\sigma ^{^{\prime }}}^{j(-)}(x_{+}-\frac{k_{2}^{+}}{eB})
\end{eqnarray}
Since the relative motion of the bound pair along the magnetic field is
nonrelativistic, we neglect $k_{3}^{\pm \text{ }}\sim me^{2}$ inside the
bispinors. The Landau level energies of the electron and positron are given
by Eqs. (\ref{eland1},\ref{eland2}).

Let us introduce the new coordinates : 
\begin{eqnarray}
Z &=&\frac{m_{n}z_{-}+m_{n^{^{\prime }}}z_{+}}{m_{n}+m_{n^{^{\prime }}}}%
,\qquad Y=\frac{1}{2}(y_{-}+y_{+}),\qquad X=\frac{1}{2}(x_{-}+x_{+}) 
\nonumber \\
z &=&z_{-}-z_{+},\qquad y=y_{-}-y_{+},\qquad x=x_{-}-x_{+}  \label{14}
\end{eqnarray}
\begin{eqnarray}
Z_{1} &=&\frac{m_{n}z_{-}^{^{\prime }}+m_{n^{^{\prime }}}z_{+}^{^{\prime }}}{%
m_{n}+m_{n^{^{\prime }}}},\qquad Y_{1}=\frac{1}{2}(y_{-}^{^{\prime
}}+y_{+}^{^{\prime }}),\qquad X_{1}=\frac{1}{2}(x_{-}^{^{\prime
}}+x_{+}^{^{\prime }})  \nonumber \\
z_{1} &=&z_{-}^{^{\prime }}-z_{+}^{^{\prime }},\qquad y_{1}=y_{-}^{^{\prime
}}-y_{+}^{^{\prime }},\qquad x_{1}=x_{-}^{^{\prime }}-x_{+}^{^{\prime }}
\label{15}
\end{eqnarray}
and new momenta : 
\begin{eqnarray}
P_{3} &=&k_{3}^{-}+k_{3}^{+},\qquad P_{2}=k_{2}^{-}+k_{2}^{+}  \nonumber \\
k_{3} &=&\frac{m_{n^{^{\prime }}}k_{3}^{-}-m_{n}k_{3}^{+}}{%
m_{n}+m_{n^{^{\prime }}}},\qquad k_{2}=\frac{1}{2}(k_{2}^{-}-k_{2}^{+})
\label{16}
\end{eqnarray}
The expression for $Q_{nn^{\prime }}$ reads now : 
\begin{eqnarray}
Q_{ilqj}^{nn^{\prime }} &=&i\int \frac{dP_{3}}{2\pi }\frac{dk_{3}}{2\pi }%
\frac{dP_{2}}{2\pi }\frac{dk_{2}}{2\pi }\frac{\exp
\{iP_{3}(Z-Z_{1})+iP_{2}(Y-Y_{1})+ik_{3}(z-z_{1})+ik_{2}(y-y_{1})\}}{%
\varepsilon -\frac{k_{3}^{2}}{2\mu _{nn^{\prime }}}-\frac{P_{3}^{2}}{%
2M_{nn^{^{\prime }}}}}  \nonumber \\
&&\times \sum_{\sigma }\sum_{\sigma ^{^{\prime }}}\Psi _{n\sigma }^{i\left(
+\right) }(X+\frac{x}{2}+\frac{P_{2}+2k_{2}}{2eB})\overline{\Psi }_{n\sigma
}^{l(+)}(X_{1}+\frac{x_{1}}{2}+\frac{P_{2}+2k_{2}}{2eB})  \nonumber \\
&&\times \Psi _{n^{^{\prime }}\sigma ^{^{\prime }}}^{q\left( -\right)
}(X_{1}-\frac{x_{1}}{2}-\frac{P_{2}-2k_{2}}{2eB})\overline{\Psi }%
_{n^{^{\prime }}\sigma ^{^{\prime }}}^{j(-)}(X-\frac{x}{2}-\frac{P_{2}-2k_{2}%
}{2eB})
\end{eqnarray}
and the positronium wave function obeys the following equation: 
\begin{eqnarray}
\hat{\Phi}\left( {\bf r},{\bf R};E\right) = &&e^{2}\int \frac{%
dZ_{1}dY_{1}dX_{1}d^{3}r_{1}}{r_{1}}\frac{dP_{3}}{2\pi }\frac{dk_{3}}{2\pi }%
\frac{dP_{2}}{2\pi }\frac{dk_{2}}{2\pi }\frac{\exp
\{iP_{3}(Z-Z_{1})+iP_{2}(Y-Y_{1})\}}{\varepsilon -\frac{k_{3}^{2}}{2\mu
_{nn^{\prime }}}-\frac{P_{3}^{2}}{2M_{nn^{^{\prime }}}}}\exp
\{ik_{3}(z-z_{1})+ik_{2}(y-y_{1})\}  \nonumber \\
&&\times \sum_{\sigma }\sum_{\sigma ^{^{\prime }}}\Psi _{n\sigma }^{i\left(
+\right) }(X+\frac{x}{2}+\frac{P_{2}+2k_{2}}{2eB})\overline{\Psi }_{n\sigma
}^{l(+)}(X_{1}+\frac{x_{1}}{2}+\frac{P_{2}+2k_{2}}{2eB})  \nonumber \\
&&\times \gamma ^{\mu }\hat{\Phi}\left( {\bf r}_{1},{\bf R}_{1};E\right)
\gamma _{\mu }\Psi _{n\sigma ^{\prime }}^{q\left( -\right) }(X_{1}-\frac{%
x_{1}}{2}-\frac{P_{2}-2k_{2}}{2eB})\overline{\Psi }_{n\sigma ^{\prime
}}^{j(-)}(X-\frac{x}{2}-\frac{P_{2}-2k_{2}}{2eB}).  \label{wfe2}
\end{eqnarray}
To separate the motion of the center of mass we subtract a bilocal phase
(Avron et al., \cite{A78}) 
\begin{equation}
{\bf PR+}e{\bf rA}({\bf R)}
\end{equation}
from $\hat{\Phi}$, with the help of the gauge choice Eq. (\ref{gauge}).
Thus, one has 
\begin{equation}
\hat{\Phi}({\bf r}_{-},{\bf r}_{+};E)={}\exp [iP_{3}^{0}Z+iP_{2}^{0}Y-ieBXy]%
\hat{\phi}({\bf r})  \label{21}
\end{equation}
By using this expression one can perform the integration on the right-hand
side of Eq. (\ref{wfe2}) over $dZ_{1}dY_{1}$. Thus, we obtain the $\delta $-
functions $\delta (P_{3}-P_{3}^{0})\delta (P_{2}-P_{2}^{0})$, which allow
for integration over $dP_{3}dP_{2}$. Therefore, we will substitute
everywhere the constants of motion $P_{3}^{0},P_{2}^{0}$ instead of $P_{3}$
and $P_{2}$ . We consider the reference frame $P_{3}^{0}=0$. Thus, the
equation for the stationary states becomes the following: 
\begin{eqnarray}
\hat{\phi}({\bf r})= &&e^{2}\int \frac{d^{3}r_{1}}{r_{1}}\frac{dk_{3}}{2\pi }%
\frac{1}{\varepsilon -\frac{k_{3}^{2}}{2\mu _{nn^{\prime }}}}\int \frac{%
dk_{2}}{2\pi }\exp \left( ik_{2}y+ieBXy\right) \int dX_{1}\exp \left(
-ik_{2}y_{1}-ieBX_{1}y_{1}\right) \exp \{ik_{3}(z-z_{1})\}  \nonumber \\
&&\times \sum_{\sigma }\sum_{\sigma ^{^{\prime }}}\Psi _{n\sigma }^{\left(
+\right) }(X+\frac{x}{2}+\frac{P_{2}^{0}+2k_{2}}{2eB})\overline{\Psi }%
_{n\sigma }^{(+)}(X_{1}+\frac{x_{1}}{2}+\frac{P_{2}^{0}+2k_{2}}{2eB}) 
\nonumber \\
&&\times \gamma ^{\mu }\hat{\phi}({\bf r}_{1})\gamma _{\mu }\Psi _{n\sigma
^{\prime }}^{\left( -\right) }(X_{1}-\frac{x_{1}}{2}-\frac{P_{2}^{0}-2k_{2}}{%
2eB})\overline{\Psi }_{n\sigma ^{\prime }}^{(-)}(X-\frac{x}{2}-\frac{%
P_{2}^{0}-2k_{2}}{2eB})  \label{22}
\end{eqnarray}
The right-hand side of Eq.(\ref{22}) depends on $z$ only through the
exponential ${}\exp (ik_{3}z)$. The action of the operator 
\begin{equation}
\widehat{H}_{z}^{0}\equiv -\frac{1}{2\mu _{nn^{^{\prime }}}}\frac{\partial
^{2}}{\ \partial z^{2}}-\varepsilon \text{.}  \label{23}
\end{equation}
on Eq.(\ref{22}) cancels the denominator on the right-hand side. Subsequent
integration over $dk_{3}$ yields $\delta (z-z_{1})$, and integration over $%
dz_{1}$ becomes trivial. Finally, we obtain : 
\begin{eqnarray}
\widehat{H}_{z}^{0}\hat{\phi}({\bf r}) &=&-e^{2}\int \frac{dx_{1}dy_{1}}{%
\sqrt{x_{1}^{2}+y_{1}^{2}+z^{2}}}\int \frac{dk_{2}}{2\pi }\exp \left(
ik_{2}y+ieBXy\right) \int dX_{1}\exp \left(
-ik_{2}y_{1}-ieBX_{1}y_{1}\right)   \nonumber \\
&&\times \sum_{\sigma }\sum_{\sigma ^{^{\prime }}}\Psi _{n\sigma }^{\left(
+\right) }(X+\frac{x}{2}+\frac{P_{2}^{0}+2k_{2}}{2eB})\overline{\Psi }%
_{n\sigma }^{(+)}(X_{1}+\frac{x_{1}}{2}+\frac{P_{2}^{0}+2k_{2}}{2eB}) 
\nonumber \\
&&\times \gamma ^{\mu }\hat{\phi}({\bf r}_{1})\gamma _{\mu }\Psi _{n\sigma
^{\prime }}^{\left( -\right) }(X_{1}-\frac{x_{1}}{2}-\frac{P_{2}^{0}-2k_{2}}{%
2eB})\overline{\Psi }_{n\sigma ^{\prime }}^{(-)}(X-\frac{x}{2}-\frac{%
P_{2}^{0}-2k_{2}}{2eB})
\end{eqnarray}
Let us now introduce the new integration variables 
\begin{equation}
\xi =X+\frac{k_{2}}{eB},\qquad \lambda =X_{1}+\frac{k_{2}}{eB}  \label{25}
\end{equation}
and denote by $x_{0}$ the distance between the orbiting centers of the
electron and positron in the plane orthogonal to the magnetic field: 
\begin{equation}
x_{0}\equiv \frac{P_{2}^{0}}{eB}
\end{equation}
This distance depends on the transverse momentum $P_{2}^{0}$, which
characterizes the motion of the mass center across the magnetic field. Then

\begin{eqnarray}
&&\widehat{H}_{z}^{0}\hat{\phi}(x,y,z)=-e^{2}eB\sum_{\sigma }\sum_{\sigma
^{^{\prime }}}\int \frac{d\xi }{2\pi }{}\exp (ieB\xi y)\Psi _{n\sigma
}^{\left( +\right) }(\xi +\frac{x+x_{0}}{2})\overline{\Psi }_{n\sigma
^{\prime }}^{(-)}(\xi -\frac{x+x_{0}}{2})  \nonumber \\
&&\times \int \frac{dx_{1}dy_{1}}{\sqrt{x_{1}^{2}+y_{1}^{2}+z^{2}}}\int
d\lambda {}\exp (-ieB\lambda y_{1}){}\overline{\Psi }_{n\sigma
}^{(+)}(\lambda +\frac{x_{1}+x_{0}}{2})\gamma ^{\mu }\hat{\phi}%
(x_{1},y_{1},z)\gamma _{\mu }\Psi _{n\sigma ^{\prime }}^{\left( -\right)
}(\lambda -\frac{x_{1}+x_{0}}{2})
\end{eqnarray}
As follows from the last equation, the wave function describing the relative
motion has the form : 
\begin{equation}
\hat{\phi}({\bf r})=\sum_{\sigma }\sum_{\sigma ^{^{\prime }}}f_{\sigma
\sigma ^{\prime }}^{n}(z)\int \frac{d\xi }{2\pi }{}\exp \left( ieB\xi
y\right) \Psi _{n\sigma }^{\left( +\right) }(\xi +\frac{x+x_{0}}{2})%
\overline{\Psi }_{n\sigma ^{^{\prime }}}^{(-)}(\xi -\frac{x+x_{0}}{2})
\end{equation}
where the functions $f_{\sigma \sigma ^{\prime }}^{n}$ depend only on the
relative coordinate $z$. These functions are solutions of the following set
of Schr\"{o}dinger-like equations: 
\begin{equation}
-\frac{1}{2\mu _{nn}}\frac{\partial ^{2}}{\partial z^{2}}f_{\sigma \sigma
^{\prime }}^{n}(z,x_{0})+\sum_{\kappa \kappa ^{\prime }}V_{\sigma \sigma ^{\prime
}}^{\kappa \kappa ^{\prime }}(z,x_{0})f_{\kappa \kappa ^{\prime
}}^{n}(z,x_{0})=\varepsilon f_{\sigma \sigma ^{\prime }}^{n}(z,x_{0})  \label{eqfnn}
\end{equation}
Here, the effective potentials represent the Coulomb interaction, averaged
over the fast transverse motion of the interacting particles : 
\begin{eqnarray}
V_{\sigma \sigma ^{\prime }}^{\kappa \kappa ^{\prime }}(z) &\equiv &-e^{2}%
\frac{eB}{2\pi }\int \frac{dxdyd\xi d\lambda {}\exp \left[ ieB\left( \xi
-\lambda \right) y\right] {}}{\sqrt{\left( x-x_{0}\right) ^{2}+y^{2}+z^{2}}}
\nonumber \\
&&\times Tr\left[ {}\overline{\Psi }_{n\sigma }^{(+)}(\lambda +\frac{x}{2}%
)\gamma ^{\mu }\Psi _{n\kappa }^{\left( +\right) }(\xi +\frac{x}{2})\right] %
\left[ \overline{\Psi }_{n\kappa ^{^{\prime }}}^{(-)}(\xi -\frac{x}{2}%
)\gamma _{\mu }\Psi _{n\sigma ^{\prime }}^{\left( -\right) }(\lambda -\frac{x%
}{2})\right] 
\end{eqnarray}
In general, the matrix $V_{\sigma \sigma ^{\prime }}^{\kappa \kappa ^{\prime
}}(z)$ consists on 16 elements corresponding to different spin orientations
of the two particles. However, in the adiabatic approximation, most of them
are zero. After integration over $d\lambda d\xi $, we obtain the effective
potentials in the following form: 
\begin{equation}
V_{\sigma \sigma ^{\prime }}^{\kappa \kappa ^{\prime }}(z)=-\frac{e^{3}B}{%
2\pi }\int \int \frac{dxdy{}\exp (-\rho ^{2}/2)}{\sqrt{%
(x-x_{0})^{2}+y^{2}+z^{2}}}R_{\sigma \sigma ^{\prime }}^{\kappa \kappa
^{\prime }}(x,y)  \label{47}
\end{equation}
where

\begin{eqnarray}
&&\ R_{\sigma \sigma ^{\prime }}^{\kappa \kappa ^{\prime }}(x,y)\equiv \exp
\left( \rho ^{2}/2\right) \int \frac{d\xi d\lambda }{2\pi }{}\exp \left[
ieB(\xi -\lambda )y\right]  \nonumber \\
&&\times \left[ {}\overline{\Psi }_{n\sigma }^{(+)}(\lambda +\frac{x}{2}%
)\gamma ^{\mu }\Psi _{n\kappa }^{\left( +\right) }(\xi +\frac{x}{2})\right] %
\left[ \overline{\Psi }_{n\kappa ^{^{\prime }}}^{(-)}(\xi -\frac{x}{2}%
)\gamma _{\mu }\Psi _{n\sigma ^{\prime }}^{\left( -\right) }(\lambda -\frac{x%
}{2})\right]
\end{eqnarray}
with $\rho =\sqrt{eB\left( x^{2}+y^{2}\right) }$.

\section{The ground band of positronium states in a superstrong magnetic
field.}

The lowest band of positronium levels corresponds to the case when both the
electron and the positron occupy the ground Landau level with $n=n^{\prime
}=0$. Only a spin combination, spin-down for the electron and spin-up for
the positron, is possible in this case. A simple calculation, using the wave
functions of Appendix A, gives the following result: 
\begin{equation}
R_{\uparrow \uparrow }^{\uparrow \uparrow }(x,y)=R_{\downarrow \downarrow
}^{\downarrow \downarrow }(x,y)=R_{\uparrow \uparrow }^{\downarrow
\downarrow }(x,y)=R_{\downarrow \downarrow }^{\uparrow \uparrow
}(x,y)=-R_{\uparrow \downarrow }^{\downarrow \uparrow }(x,y)=-R_{\downarrow
\uparrow }^{\uparrow \downarrow }(x,y)=R_{\uparrow \downarrow }^{\uparrow
\downarrow }(x,y)=0
\end{equation}
and 
\begin{equation}
R_{\downarrow \uparrow }^{\downarrow \uparrow }(x,y)=1.
\end{equation}
Consequently, the wave functions of the lowest band, which will be labeled
by the quantum number $l$, have the following form : 
\begin{equation}
\hat{\phi}_{00l}^{P_{2}^{0}}({\bf r})=f_{\downarrow \uparrow
}^{(0,l)}(z,x_{0})\int \frac{d\xi }{2\pi }{}\exp (ieB\xi y)\Psi
_{0\downarrow }^{\left( +\right) }(\xi +\frac{x+x_{0}}{2})\overline{\Psi }%
_{0\uparrow }^{(-)}(\xi -\frac{x+x_{0}}{2})  \label{52}
\end{equation}
Using the formulae for bispinors corresponding to Landau states (see
Appendix A), we can integrate this expression over $d\xi $. The result is : 
\begin{equation}
\hat{\phi}_{00l}^{P_{2}^{0}}({\bf r})=f_{\downarrow \uparrow
}^{(0,l)}(z,x_{0})\sqrt{\frac{eB}{2\pi }}{}\exp \{-\frac{eB}{4}%
[(x+x_{0})^{2}+y^{2}]\}\left( 
\begin{tabular}{l}
$0$ \\ 
$1$ \\ 
$0$ \\ 
$0$%
\end{tabular}
\right) \left( 
\begin{array}{cccc}
0 & 0 & 0 & 1
\end{array}
\right)   \label{53}
\end{equation}
The wave function for the relative motion along the magnetic field can be
found by solving the following Schr\"{o}dinger equation : 
\begin{equation}
-\frac{1}{2\mu _{00}}\frac{\partial ^{2}}{\partial z^{2}}f_{\downarrow
\uparrow }^{(0,l)}(z,x_{0})+V_{\downarrow \uparrow }^{\downarrow \uparrow
}(z,x_{0})f_{\downarrow \uparrow }^{(0,l)}(z,x_{0})=\varepsilon
_{00}^{l}\left( x_{0}\right) f_{\downarrow \uparrow }^{(0,l)}(z,x_{0})
\label{greq}
\end{equation}
with $\mu _{00}=m/2$. The effective potential is 
\begin{equation}
V_{00}(z,x_{0})\equiv V_{\downarrow \uparrow }^{\downarrow \uparrow
}(z,x_{0})=-\frac{2e^{2}}{\pi }\int_{0}^{\infty }\frac{\rho d\rho {}\exp
(-\rho ^{2}/2)}{\sqrt{(\rho a_{L}-x_{0})^{2}+z^{2}}}\ {\sf K}\left( i\sqrt{%
\frac{4\rho x_{0}a_{L}}{(\rho a_{L}-x_{0})^{2}+z^{2}}}\right)   \label{54}
\end{equation}
and 
\begin{equation}
{\sf K}\left( k\right) =\int_{0}^{\pi /2}\frac{d\theta }{\sqrt{1-k^{2}\sin
^{2}\theta }}  \label{elliptic}
\end{equation}
is the elliptic integral of the first kind.

Eq. (\ref{54}) depends on the parameter $x_{0}=P_{2}^{0}/eB$, therefore the
eigen-functions $f_{\downarrow \uparrow }^{(0,l)}$, as well as the discrete
spectrum eigen-energies $\varepsilon _{00}^{l}$, depend on the quantity $%
P_{2}^{0}$ , which characterizes the relativistic motion of the center of
mass across the magnetic field. Unfortunately the integration in Eq. (\ref
{54}) can not be done analytically. By this reason, we consider several
limiting cases.

\subsection{The case of small $x_0$ .}

We first consider the case of a small distance between the centers of Landau
orbits in the plane orthogonal to the magnetic field: 
\begin{equation}
\sqrt{eB}x_{0}=\frac{P_{2}^{0}}{\sqrt{eB}}\ll 1  \label{55}
\end{equation}
This yields: 
\begin{equation}
V_{00}(z)\approx -e^{2}\sqrt{\frac{\pi eB}{2}}\left[ 1-{}%
\mathop{\rm erf}%
\left( |z|\sqrt{\frac{eB}{2}}\right) \right] {}\exp \left( \frac{eBz^{2}}{2}%
\right)  \label{56}
\end{equation}
with 
\begin{equation}
\mathop{\rm erf}%
(x)=\frac{2}{\sqrt{\pi }}\int_{0}^{x}dt\exp \left( -t^{2}\right)
\end{equation}
being the error function. At distances $\left| z\right| \gg a_{L}$, the
potential behaves as the one-dimensional Coulomb potential 
\begin{equation}
V_{00}(z)\approx -\frac{e^{2}}{\left| z\right| }  \label{Coul}
\end{equation}
while for $\left| z\right| \ll a_{L}$ goes to a constant 
\begin{equation}
V_{00}(0)=-\sqrt{\frac{\pi }{2}}\frac{e^{2}}{a_{L}}
\end{equation}
(This result differs from that given by Usov and Shabad by the factor 
$\sqrt{\pi /2}$).

For shallow levels, the characteristic size of the atom along the $Oz$ axis
is of the order $a_{B}\gg a_{L\text{ }}$, so that the potential can be
considered to have the form Eq. (\ref{Coul}), and the energy levels should
be close to Balmer's spectrum. This statement is not valid for deep levels.
Eq. (\ref{greq}), with the potential Eq. (\ref{56}), can be solved
numerically. However, an analytical estimate is of interest. To obtain such
an estimate, we replace the potential Eq. (\ref{56}) by a simple function of
the form 
\begin{equation}
V_{00}(z)\approx -\frac{e^{2}}{\left| z\right| +\sqrt{\frac{2}{\pi }}a_{L}}
\end{equation}
which slightly differs from Eq. (\ref{56}) only in the region $\left|
z\right| \sim a_{L}$, and has the same asymptotic forms.

\subsubsection{Discrete spectrum.}

Solutions to the Schr\"{o}dinger equation with this potential have been
investigated by Loudon \cite{Loudon} for negative energy. For completeness,
we quote shortly this calculations, which will be used later for
investigation of the states of positive energy. We define, as in \cite
{Loudon}, a dimensionless quantity $\alpha $ by writing 
\begin{equation}
\varepsilon =-\frac{1}{2\mu _{00}a_{00}^{2}\alpha ^{2}}  \label{alpha}
\end{equation}
with $a_{00}=\left( \mu _{00}e^{2}\right) ^{-1}=2a_{B}$, and replace the
independent variable $z$ in Eq. (\ref{greq}) by 
\begin{eqnarray}
z^{\prime } &=&\frac{2}{\alpha a_{00}}\left( \sqrt{\frac{2}{\pi }}%
a_{L}+z\right) ,\hspace{1in}\text{for }z>0  \nonumber \\
z^{\prime } &=&-\frac{2}{\alpha a_{00}}\left( \sqrt{\frac{2}{\pi }}%
a_{L}-z\right) ,\hspace{1in}\text{for }z<0
\end{eqnarray}
whereupon, for $z\neq 0$, the equation takes the Whittaker's form of the
confluent hypergeometric equation 
\begin{equation}
\frac{d^{2}f}{dz^{\prime 2}}-\frac{1}{4}f+\frac{\alpha }{\left| z^{\prime
}\right| }f=0  \label{whitf}
\end{equation}
This equation has two independent solutions, given by Whittaker's functions.
The first of them 
\begin{equation}
M_{\alpha ,\frac{1}{2}}\left( z^{\prime }\right) =\exp \left( -z^{\prime
}/2\right) z^{\prime }\Phi (1-\alpha ,2;z^{\prime })
\end{equation}
where $\Phi (1-\alpha ,2;z^{\prime })$ is the confluent hypergeometric
function of the first kind, diverges as $\left( z^{\prime }\right) ^{-\alpha
}\exp \left( z^{\prime }/2\right) $ for large $z$. Since for a bound state
any solution of this equation must go to zero when $\left| z^{\prime
}\right| $ tends to infinity, we choose the second solution, which for $z>0$
is 
\begin{equation}
W_{\alpha ,\frac{1}{2}}\left( z^{\prime }\right) =\exp \left( -z^{\prime
}/2\right) z^{\prime }\Psi (1-\alpha ,2;z^{\prime })
\end{equation}
where $\Psi (1-\alpha ,2;x)$ is the confluent hypergeometric function of the
second kind : 
\begin{equation}
\Psi (1-\alpha ,2;x)=\frac{1}{\Gamma \left( -\alpha \right) }\{\,\Phi
(1-\alpha ,2;x)[\log x+\psi \left( 1-\alpha \right) -\psi \left( 1\right)
-\psi \left( 2\right) ]-\frac{1}{\alpha x}+\sum_{r=1}^{\infty }\frac{\left(
1-\alpha \right) _{r}}{r!\left( r+1\right) !}A_{r}x^{r}\}  \label{whitteker}
\end{equation}
Here 
\begin{equation}
A_{r}=\sum_{n=0}^{r-1}\left[ \frac{1}{n+1-\alpha }-\frac{1}{n+1}-\frac{1}{n+2%
}\right] ,\hspace{1in}\left( c\right) _{r}=\frac{\Gamma \left( c+r\right) }{%
\Gamma \left( c\right) }
\end{equation}
The function $\psi \left( s\right) =\Gamma ^{\prime }\left( s\right) /\Gamma
\left( s\right) $ is the logarithmic derivative of the gamma function. The
solutions for positive and negative $z$ can be joined together to form
either even or odd wave-functions. For an odd state we require 
\begin{equation}
W_{\alpha ,\frac{1}{2}}\left( \sqrt{\frac{2}{\pi }}\frac{2a_{L}}{\alpha
a_{00}}\right) =0  \label{negat}
\end{equation}
while for an even state 
\begin{equation}
\left[ \frac{d}{dz^{\prime }}W_{\alpha ,\frac{1}{2}}\left( z^{\prime
}\right) \right] _{z^{\prime }=\sqrt{\frac{2}{\pi }}\frac{2a_{L}}{\alpha
a_{00}}}=0  \label{posit}
\end{equation}

To find solutions to Eqs. (\ref{negat}) and (\ref{posit}), which give the
eigenvalues of the system, we keep only terms which are dominant when 
\begin{equation}
x=\sqrt{\frac{2}{\pi }}\frac{2a_{L}}{\alpha a_{00}}
\end{equation}
is very small, and $\alpha $ is close to a positive integer. The eigenvalue
conditions then become :

Odd state: 
\begin{equation}
\psi \left( 1-\alpha \right) \sqrt{\frac{2}{\pi }}\frac{2a_{L}}{\alpha a_{00}%
}-\frac{1}{\alpha }=0
\end{equation}

Even state: 
\begin{equation}
\log \frac{2}{\alpha a_{00}}\sqrt{\frac{2}{\pi }}a_{L}+\,\psi \left(
1-\alpha \right) =0  \label{77}
\end{equation}
Assuming $\alpha \rightarrow l$ where $l=1,2,...$ we can replace 
\begin{equation}
\,\psi \left( 1-\alpha \right) \rightarrow \frac{1}{\alpha -l}
\end{equation}
Then, the quantum defects $\delta _{l}=\alpha -l$ are given by

Odd state: 
\begin{equation}
\delta _l=\frac 2{a_{00}}\sqrt{\frac 2\pi }a_L  \label{quantodd}
\end{equation}

Even state: 
\begin{equation}
\delta _l=-\left[ \log \frac 2{la_{00}}\sqrt{\frac 2\pi }a_L\right] ^{-1}
\label{quanteven}
\end{equation}
For the adiabatic case one has $\delta _l\ll 1$ if $\log (a_{00}/a_L)\gg 1$.

In addition to this series of states having their quantum numbers $\alpha $
close to positive integers, there is another state having $\alpha $ close to
zero. For such a value of $\alpha $, $\psi \left( 1-\alpha \right) $ is no
longer an important term in Eq. (\ref{77}), and $1/\alpha $ becomes the
dominant term in $\alpha $. In this case, the equation for the eigenvalues
becomes 
\begin{equation}
\log \frac{2}{\alpha a_{00}}\sqrt{\frac{2}{\pi }}a_{L}=-\frac{1}{2\alpha }
\label{quanteven0}
\end{equation}
The quantum defect $\delta _{l}=\alpha $ for the state $l=0$ can be found
from Eq. (\ref{quanteven0}) by iteration. To the first order in $a_{L}/a_{00}
$ its binding energy is given by 
\begin{equation}
\varepsilon _{1}^{\left( 0\right) }=\varepsilon _{2}^{\left( 0\right)
}\approx -me^{4}\left( \log \sqrt{\frac{\pi }{2}}\frac{a_{00}}{2a_{L}}%
\right) ^{2}
\end{equation}
To this state there corresponds the even wave-function 
\begin{equation}
f_{\downarrow \uparrow }^{(0,0)}(z,0)=\left[ \frac{1}{a_{00}}\ln \left( 
\sqrt{\frac{\pi }{2}}\frac{a_{00}}{a_{L}}\right) \right] ^{1/2}\exp \left[ -%
\frac{\left| z\right| }{a_{00}}\ln \left( \sqrt{\frac{\pi }{2}}\frac{a_{00}}{%
a_{L}}\right) \right] 
\end{equation}

\subsubsection{Continuous spectrum.}

The spectrum for positive energies is continuous. Now, the variable $%
z^{\prime }$, as well as the $\alpha $ value defined by Eq. (\ref{alpha})
are imaginary quantities. Let us write them as 
\begin{equation}
\alpha =\frac{i}{ka_{00}}
\end{equation}
with $k=\sqrt{2\mu _{00}\varepsilon }$, and define: 
\begin{eqnarray}
z^{\prime } &=&-2ik\left( \sqrt{\frac{2}{\pi }}a_{L}+z\right) ,\hspace{1in}%
\text{for }z>0  \nonumber \\
z^{\prime } &=&2ik\left( \sqrt{\frac{2}{\pi }}a_{L}-z\right) ,\hspace{1in}%
\text{for }z<0
\end{eqnarray}
In the region $z^{\prime }>0$, Eq. (\ref{whitf}) has two
linearly-independent solutions, given by the Whittaker's functions 
\begin{equation}
f_{1}^{(k)}(z)=C_{1}W_{\frac{i}{ka_{00}},\frac{1}{2}}\left( -2ik\left[ \sqrt{%
\frac{2}{\pi }}a_{L}+z\right] \right)   \nonumber
\end{equation}
\begin{equation}
f_{2}^{(k)}(z)=C_{2}M_{\frac{i}{ka_{00}},\frac{1}{2}}\left( -2ik\left[ \sqrt{%
\frac{2}{\pi }}a_{L}+z\right] \right) 
\end{equation}
where $C_{1}$ and $C_{2}$ are constants, which should be determined by the
normalization conditions. To find these constants, let us consider the
asymptotic behavior of Whittaker's functions when $z\rightarrow \infty $%
\begin{equation}
W_{\frac{i}{ka_{00}},\frac{1}{2}}\left( -2ik\left[ \sqrt{\frac{2}{\pi }}%
a_{L}+z\right] \right) \simeq \exp \left( -\frac{3\pi }{2ka_{00}}\right)
\exp \left( ikz+\frac{i}{ka_{00}}\ln 2kz+i\sqrt{\frac{2}{\pi }}ka_{L}\right) 
\end{equation}
\begin{equation}
M_{\frac{i}{ka_{00}},\frac{1}{2}}\left( -2ik\left[ \sqrt{\frac{2}{\pi }}%
a_{L}+z\right] \right) \simeq \frac{ka_{00}\exp \left( \frac{3\pi }{2ka_{00}}%
\right) }{\left| \Gamma \left( \frac{i}{ka_{00}}\right) \right| }\exp \left(
-ikz-\frac{i}{ka_{00}}\ln 2kz-i\sqrt{\frac{2}{\pi }}ka_{L}+i\frac{3\pi }{2}%
-i\delta _{k}\right) 
\end{equation}
where 
\begin{equation}
\delta _{k}=\arg \Gamma \left( \frac{i}{ka_{00}}\right) 
\end{equation}
If we choose to normalize the eigen-functions in the following way : 
\begin{equation}
\int_{-\infty }^{\infty }f^{(k)}(z)f^{(k^{\prime })}(z)dz=2\pi \delta \left(
k-k^{\prime }\right) 
\end{equation}
then the normalization factors are 
\begin{equation}
C_{1}=\exp \left( \frac{3\pi }{2ka_{00}}\right) ,\qquad C_{2}=\frac{1}{%
ka_{00}}\left| \Gamma \left( \frac{i}{ka_{00}}\right) \right| \exp \left( -%
\frac{3\pi }{2ka_{00}}\right) 
\end{equation}
Indeed, the asymptotic form of the wave functions in this case are in a good
agreement with the general form of the normalized one-dimensional wave
functions for a continuous spectrum, in the form $\exp \left( \pm ikz\right) 
$. The logarithmic term in the exponentially grows much slower than $z$.
This term is not important when calculating the normalization integral,
which diverges at infinity.

\subsection{The case of large $x_0$.}

The bound pair of an electron and positron with a small magnitude of $%
P_{2}^{0}$ in a superstrong magnetic field is only of academic interest. The
known mechanism of positronium production in a pulsar magnetosphere \cite
{LO85b}, \cite{ShU85}, \cite{HRW85} assumes that the created bound pair has
a quasi momentum $P_{2}^{0}\gtrsim 2m$. For a typical pulsar magnetic field $%
B\lesssim 0.1B_{0}$, this corresponds to the opposite limit : 
\begin{equation}
\sqrt{eB}x_{0}=\frac{P_{2}^{0}}{\sqrt{eB}}\gg 1  \label{57}
\end{equation}
In this limit, the effective potential \ Eq. (\ref{54}) takes the simple
analytical form 
\begin{equation}
V_{00}(z/x_{0})\approx -\frac{e^{2}}{x_{0}}\frac{1}{\sqrt{1+z^{2}/x_{0}^{2}}}
\label{58}
\end{equation}
For $P_{2}^{0}\gtrsim 2m$ and $B=0.1B_{0}$, i.e. $\sqrt{eB}x_{0}\gtrsim 6$, 
the curve given by Eq.(\ref{58}) practically coincides with the exact
potential. In Fig. 1 we show the potential, in units of $me^{4}$, versus the
dimensionless distance $me^{2}z$, for two values of $P_{2}^{0}$. We also
show, for comparison, the approximation 
\begin{equation}
V_{00}(z)\approx -\frac{e^{2}}{\left| z\right| +x_{0}}
\label{USapprox}
\end{equation}
used by Usov and Shabad.

In spite of the relatively simple expression given by Eq. (\ref{58}), there
are not available analytical solutions of the Schr\"{o}dinger problem for
stationary states in this potential. For this reason, we have made a
numerical calculation in order to determine the energy $\varepsilon _{00}^{l}
$ for bound states. The results, for the first five levels, are shown in
Table 1, where three values of the dimensionless parameter $\xi _{0}$,
defined as 
\begin{equation}
\xi _{0}\equiv \frac{x_{0}}{a_{00}},
\end{equation}

were considered. 
For large values $\xi _{0}\gg 1$, these results can be approximated by the
following formula: 
\begin{equation}
\frac{\varepsilon _{00}^{l}}{me^{4}}=-\frac{\left[ \sqrt{(l+1)^{2}+8\xi _{0}}%
-(l+1)\right] ^{2}}{16\xi _{0}^{2}}
\end{equation}

\section{Conclusions}

We have investigated the bound states of an electron and positron in the
presence of a superstrong magnetic field ($B>>10^{9}G$), as it can appear in
some neutron stars. We have given a completely relativistic description of
the positronium motion across the magnetic field. Our starting point is the
Bethe-Salpeter equation for the positron and the electron at the lowest
order in the electromagnetic coupling constant. The effects of the strong
external magnetic field are incorporated through the exact solutions of the
Dirac equation for the interacting particles. The Bethe-Salpeter equation
then involves a summation over all possible Landau levels of those
particles. As we have shown, however, this equation can be transformed into
a set of coupled Schr\"{o}dinger-like equations under the hypothesis of the
so-called adiabatic approximation, which is valid for superstrong magnetic
fields. In this case, only one Landau level per interacting particle is
relevant.

We have concentrated ourselves to some particular cases of particular
interest, like the ground band of positronium, where we found some
differences with previous results.

{\bf Acknowledgments}

This work has been partially supported by the Spanish Grants DGES PB97-1432
and AEN99-0692. L. B. Leinson would like to thank the Russian Foundation for
Fundamental Research, Grant 00-02-16271.

\appendix

\section{Solutions of the Dirac equation in a constant magnetic field}

The wave functions we use are solutions of the Dirac equation in a constant
magnetic field directed along the $z$-axis. By the use of the asymmetric
(Landau) gauge Eq. (\ref{gauge}), these four wave- functions, with definite
third-component spin direction, can be expressed in terms of the stationary
wave functions as: 
\begin{equation}
\frac{1}{\sqrt{L_{2}L_{3}}}\exp (-i\epsilon E_{n}t)\exp \left[ i\epsilon
\left( k_{3}z+k_{2}y\right) \right] \Psi _{n\sigma }^{\epsilon }(\xi
_{\epsilon },k_{3})  \label{Eq.2A}
\end{equation}
The wave functions are normalized in a volume $\left( L_{1}L_{2}L_{3}\right) 
$ , $n$ is the quantum number of the Landau level, $\sigma =\uparrow
,\downarrow $ is the spin projection along the $z$-axis, $\epsilon =+1$ ($%
\epsilon =-1$) indicates the electron (positron) states, and $k_{2},k_{3}$
are the momenta in the $y$ and $z$ directions, respectively. The functions $%
\Psi _{n\sigma }^{\epsilon }(\xi _{\epsilon },k_{3})$, with 
\begin{equation}
\xi _{\pm }=\sqrt{eB}\left( x\pm \frac{k_{2}}{eB}\right)  \label{Eq.3A}
\end{equation}
are given by 
\begin{equation}
\Psi _{n\downarrow }^{+}=C_{n}^{1/2}\left( 
\begin{tabular}{l}
$-ik_{3}\sqrt{E_{n}^{0}-m}\varphi _{n-1}(\xi _{+})$ \\ 
$(E_{n}+E_{n}^{0})\sqrt{E_{n}^{0}+m}\varphi _{n}(\xi _{+})$ \\ 
$-i(E_{n}+E_{n}^{0})\sqrt{E_{n}^{0}-m}\varphi _{n-1}(\xi _{+})$ \\ 
$-k_{3}\sqrt{E_{n}^{0}+m}\varphi _{n}(\xi _{+})$%
\end{tabular}
\right)  \label{Eq.4A}
\end{equation}
\begin{equation}
\Psi _{n\uparrow }^{+}=C_{n}^{1/2}\left( 
\begin{tabular}{l}
$(E_{n}+E_{n}^{0})\sqrt{E_{n}^{0}+m}\varphi _{n-1}(\xi _{+})$ \\ 
$-ik_{3}\sqrt{E_{n}^{0}-m}\varphi _{n}(\xi _{+})$ \\ 
$k_{3}\sqrt{E_{n}^{0}+m}\varphi _{n-1}(\xi _{+})$ \\ 
$i(E_{n}+E_{n}^{0})\sqrt{E_{n}^{0}-m}\varphi _{n}(\xi _{+})$%
\end{tabular}
\right)  \label{Eq.5A}
\end{equation}
\begin{equation}
\Psi _{n\downarrow }^{-}=C_{n}^{1/2}\left( 
\begin{tabular}{l}
$k_{3}\sqrt{E_{n}^{0}+m}\varphi _{n-1}(\xi _{-})$ \\ 
$-i(E_{n}+E_{n}^{0})\sqrt{E_{n}^{0}-m}\varphi _{n}(\xi _{-})$ \\ 
$(E_{n}+E_{n}^{0})\sqrt{E_{n}^{0}+m}\varphi _{n-1}(\xi _{-})$ \\ 
$ik_{3}\sqrt{E_{n}^{0}-m}\varphi _{n}(\xi _{-})$%
\end{tabular}
\right)  \label{Eq.6A}
\end{equation}
\begin{equation}
\Psi _{n\downarrow }^{-}=C_{n}^{1/2}\left( 
\begin{tabular}{l}
$i(E_{n}+E_{n}^{0})\sqrt{E_{n}^{0}-m}\varphi _{n-1}(\xi _{-})$ \\ 
$-k_{3}\sqrt{E_{n}^{0}+m}\varphi _{n}(\xi _{-})$ \\ 
$ik_{3}\sqrt{E_{n}^{0}-m}\varphi _{n-1}(\xi _{-})$ \\ 
$(E_{n}+E_{n}^{0})\sqrt{E_{n}^{0}+m}\varphi _{n}(\xi _{-})$%
\end{tabular}
\right)  \label{Eq.7A}
\end{equation}
with 
\begin{equation}
C_{n=}\frac{1}{4E_{n}E_{n}^{0}(E_{n}+E_{n}^{0})}
\end{equation}
\begin{equation}
E_{n}=\sqrt{m^{2}+2neB+k_{3}^{2}}
\end{equation}
\begin{equation}
E_{n}^{0}=\sqrt{m^{2}+2neB}\equiv m_{n}
\end{equation}
The functions $\varphi _{n}(\xi )$ are the eigenfunctions of the
one-dimensional harmonic oscillator, normalized with respect to $x$.

\newpage

\begin{table*}[hbt]
\caption{Bounding energy of the first positronium energy levels, for
different values of the parameter $\xi _{0}$}
\label{tab1}
\begin{tabular}{|l|l|l|l|l|l|}
\hline
$\xi _{0}$ & $l=0$ & $l=1$ & $l=2$ & $l=3$ & $l=4$ \\ \hline
0.1 & -1.9093 & -0.2371 & -0.1283 & -0.0609 & -0.0433 \\ \hline
1 & -0.3349 & -0.1374 & -0.0757 & -0.0463 & -0.0318 \\ \hline
10 & -0.0432 & -0.0319 & -0.0240 & -0.0186 & -0.0147 \\ \hline
\end{tabular}
\end{table*}

\vspace{2cm}

\begin{figure}
\includegraphics[width=10cm,angle=270]{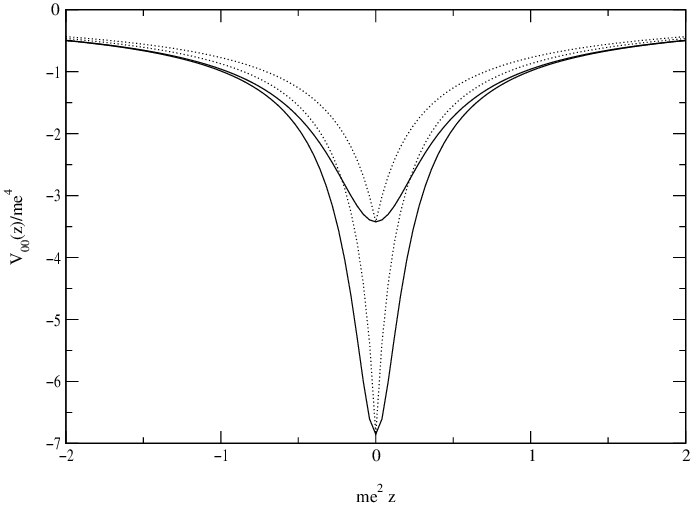}
\caption{Interaction potential of positronium, in units of $m e^4$,
as given by our Eq. (\ref{58})
(solid lines). We also show for comparison the approximation, Eq.
(\ref{USapprox})
used by Usov and Shabad (dotted lines). In both cases, the deepest curve
corresponds to $P_{2}^{0}=2m$, and the other one to $P_{2}^{0}=4m$.}
\label{Fig. 1}
\end{figure}

\end{document}